# Ultra-wide-band slow light in photonic crystal coupled-cavity waveguides

Yiming Lai[1,2], Mohamed Sabry Mohamed[3], Boshen Gao[4], Momchil Minkov[3], Robert W. Boyd[1], Vincenzo Savona[3], Romuald Houdré[3], and Antonio Badolato[1,2]

[1]*Department of Physics University of Ottawa, Ottawa, Ontario K1N 6N5, Canada*
[2]*Center for Nanoscale Science and Technology, National Institute of Standards and Technology, Gaithersburg, MD 20899, U.S.A.*
[3]*Institut de Physique, Ecole Polytechnique Fédérale de Lausanne (EPFL), CH-1015 Lausanne, Switzerland*
[4]*The Institute of Optics, University of Rochester, Rochester, NY 14627, U.S.A.*

Slow light propagation in structured materials is a highly promising approach for realizing on-chip integrated photonic devices based on enhanced optical nonlinearities[1,2]. One of the most successful research avenues consists in engineering the band dispersion of light-guiding photonic crystal (PC) structures[3,4]. The primary goal of such devices is to achieve slow-light operation over the largest possible bandwidth, with large group index, minimal index dispersion, and constant transmission spectrum. Such features are required to enable multimode or pulsed operation as they suppress pulse distortion and the onset of echoes. A commonly adopted figure of merit for this set of features is the group-index-bandwidth product (GBP). Significant progress in recent years has brought to photonic structures with increasingly high GBP values[5].

Here, we report on the experimental demonstration of to date record high GBP in silicon-based coupled-cavity waveguides (CCWs)[6,7,8,9,10,11] operating at telecom wavelengths. Our results rely on novel CCW designs, optimized using a genetic algorithm, and refined nanofabrication processes[12]. The schematic design of our CCW unit cell is shown in Fig. 1a. It comprises two L3 photonic crystal nanocavities[13] (PCNs) separated by $5a$ in the *x*-direction and $a\sqrt{3}$ in the propagation direction *y* (where *a* is the lattice constant of the PC). Defining the CCW flat-band operation bandwidth as the spectral range $\Delta\omega$ where the group index ($n_g = c\, dk/d\omega$) deviates from a mean value $\langle n_g \rangle$ by less than[14] ±10%, we maximized the group-index bandwidth product (GBP) (= $\langle n_g \rangle \Delta\omega/\omega$ for $n_g \sim \langle n_g \rangle$), while concomitantly minimizing losses, through the combination of three criteria. First, the shortest possible spatial period results in a large Brillouin zone and thus in a large GBP for a given bandwidth. Second, the small first-neighbor coupling, stemming from the staggered geometry, minimizes the overall bandwidth of the guided band, as suggested by a tight-binding (TB) description, thus increases the GBP. Third, the maximization of the quality factor ($Q$) of each single L3 PCCs [15] minimizes the intrinsic losses of the CCW. We carried out the optimization through a genetic algorithm[16,17] combined with the guided mode expansion (GME) method[18]. As an objective function, we choose the GBP with an additional price if the maximum radiation loss per unit time ($L_e$) of the electric field intensity exceeded $L_e$ = 100 dB/ns. To maximize the objective function, we introduced four free parameters (Fig. 1a, $\Delta r_{1,2,3}$ and $\Delta x$) that were varied simultaneously. Radii and positions of the blue air holes ($\Delta r_3$ and $\Delta x$ respectively) mostly affected the $Q$ of each L3 PCCs. Radii of red and green air holes ($\Delta r_1$ and $\Delta r_2$ respectively) instead affected primarily the first- and second-neighbor couplings respectively. Fig. 1b-c show the computed band structure of a genetically optimized CCW and the corresponding $n_g$. Our numerical simulations demonstrated a GBP value of 0.47, over a 19.5 nm bandwidth, for the set of parameters ($\Delta r_1$, $\Delta r_2$, $\Delta r_3$, $\Delta x$) = (- 0.0385a, - 0.0279a, - 0.0759a, 0.1642a). Higher theoretical GPB values of up to 0.66 were obtained at the expense of a narrower bandwidth (see Supplementary Section S3). To achieve CCW performance with a flat-band centered near 1550 nm and with standard silicon membrane thickness of $d$ = 220 nm (see Methods), we locked the lattice constant $a$ = 400 nm and bulk air hole radius $r$ = 0.25$a$. Fig. 1d shows the top view scanning electron microscope image of one of our CCWs composed of 50 PCCs (50-CCW) fabricated in silicon-on-insulator (see Methods). As detailed in the Supplementary Section S1, all the CCWs were coupled to two separate input-output PC waveguides (red arrows in Fig. 1d) each of which (not shown in the figure) was butt-coupled to silicon-strip waveguides and then to spot size converters for efficient end-fire coupling. A nano-tether-based structure surrounding the device (not shown in figure) was engineered to achieve buckling-free suspended membranes[19].

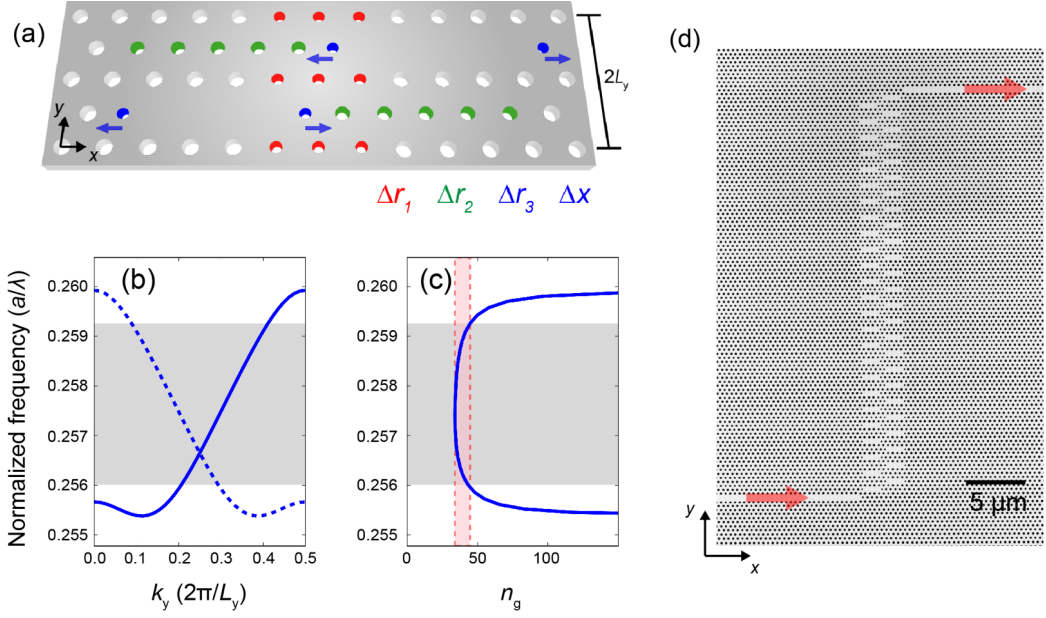

**Figure 1** | (a) Schematic of the CCW unit cell. The radii of the colored holes are modified ($\Delta r_{1,2,3}$) with respect to the PC bulk holes. The blue holes are also shifted outward ($\Delta x$). (b) GME-simulated band structure and (c) corresponding $n_g$ in normalized frequency units ($a/\lambda = \omega a/2\pi c$). The pink region indicates the operation bandwidth. (d) SEM top view image of a 50-CCW. Red arrows indicate the light input and output.

To study the transmission of the designed CCWs with respect to the number of PCCs composing the CCWs, we fabricated five groups of CCWs formed by 50, 100, 200, 400, and 800 total PCCs. Fig. 2a shows the typical transmission profiles of those groups over their operation bandwidth. The transmission remained flat with variation < 10 dB over a wide-band for CCWs containing up to 400 PCCs (400-CCW). While for the 800-CCW, the transmission started dropping as extrinsic optical scattering losses (likely due to structural disorder and material absorption) became significant[20], as also hinted by the more pronounced transmission fluctuations[21]. In general, for all CCW groups, part of the transmission fluctuations, especially at the extrema of the slow light band, display regularities and are thus likely caused by the abrupt coupling of the CCW to the input and output waveguides. A proper apodization of the design[14] may prevent this effect and still improve the performance of the device. Fig. 2b shows the real-space image of an 800-CCW as light propagates along the waveguide. The propagation loss of the CCW was estimated first by filtering out the strong scattering at the input edge of the CCW and then by integrating the real space profile along the $x$-direction. From the resulting intensity versus $y$-position, we fitted the loss per unit length[22]. By combining it with the measured $n_g$ (discussed later), the radiation loss per unit time at the center of the transmission band (1562.5 nm) was determined to be $L_e \sim 56$ dB/ns.

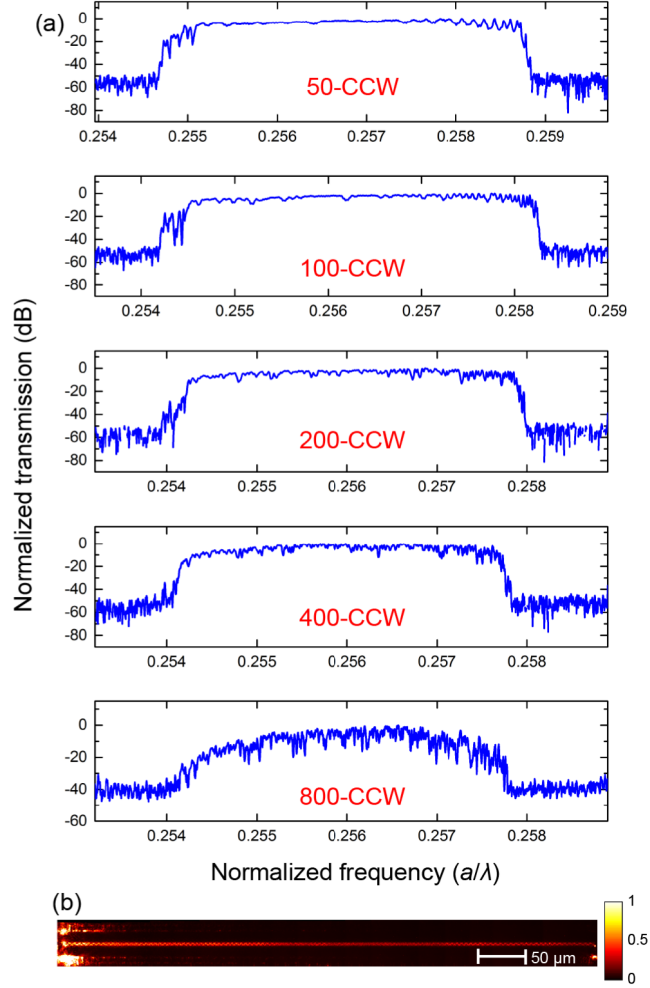

**Figure 2 |** (a) Normalized transmission spectrum of CCWs made up of 50, 100, 200, 400, and 800 PCCs, respectively. (b) Optical image of an 800-CCW when light is propagating from left to right.

To comprehensively investigate our different CCWs we applied two complementary techniques: Fourier-space imaging (FSI)[23] and Mach-Zehnder (MZ) interferometry[24]. For FSI, the out-of-plane emission of the coupled PCCs is linked to the wave vector of the propagating mode through momentum conservation at the silicon/air interface. By measuring the far-field radiation angle, the in-plane Bloch wave vector $k_y$ of the mode is determined, and the dispersion relation can be reconstructed by scanning the light frequency. Fig. 3a shows the measured dispersion relation for a 50-CCW and the excellent agreement with the simulated dispersion relation using the GME method (dashed white curve). From the measured dispersion relation, we obtained by numerical differentiation of the peak intensities the $n_g$ of the CCW (red crosses in Fig. 3b), which are compared with the $n_g$ curves obtained from the TB model (blue curve) and the GME method (dashed black curve). By fitting the experimental data with the TB model, we determined $\langle n_g \rangle = 41.0$ and $\Delta\lambda = 17.7 \pm 0.5$ nm, corresponding to GBP = $0.47 \pm 0.01$ (with the error accounting for the uncertainty in fitting the dispersion curves). The prediction of the GME model for the same device gave $\langle n_g \rangle = 37$ over an operational bandwidth of $\Delta\lambda = 19.5$ nm. To our knowledge, this is the highest experimental GBP ever reported in PC-based slow light devices. It is also worth noting the excellent agreement between experimental data and theory, with the measured GBP being nearly the same in all fabricated devices of nominally the same design (with the highest GBP observed in the 200-CCW). To explore the space of parameters around this optimal design we fabricated three additional series of CCW devices. For these devices, we measured a higher group index, at the expense of consistently lower GBP values, as detailed in the Supplementary Section S3.

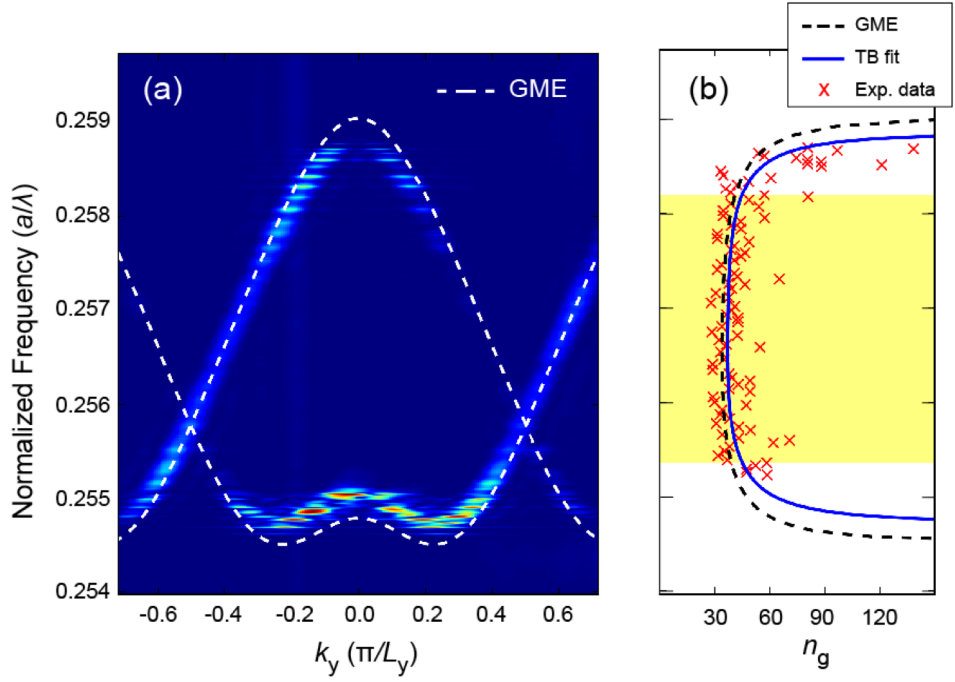

**Figure 3** | (a) Photonic band structure of a CCWs made up of 50 PCCS (50-CCW) measured by FSI overlaid with the GME simulation (white dashed line). (b) Red crosses: experimental $n_g$ calculated from data in (b). Blue continuous curve: fit of the experimental $n_g$ using the TB model. Black dashed curve: prediction of the GME simulation.

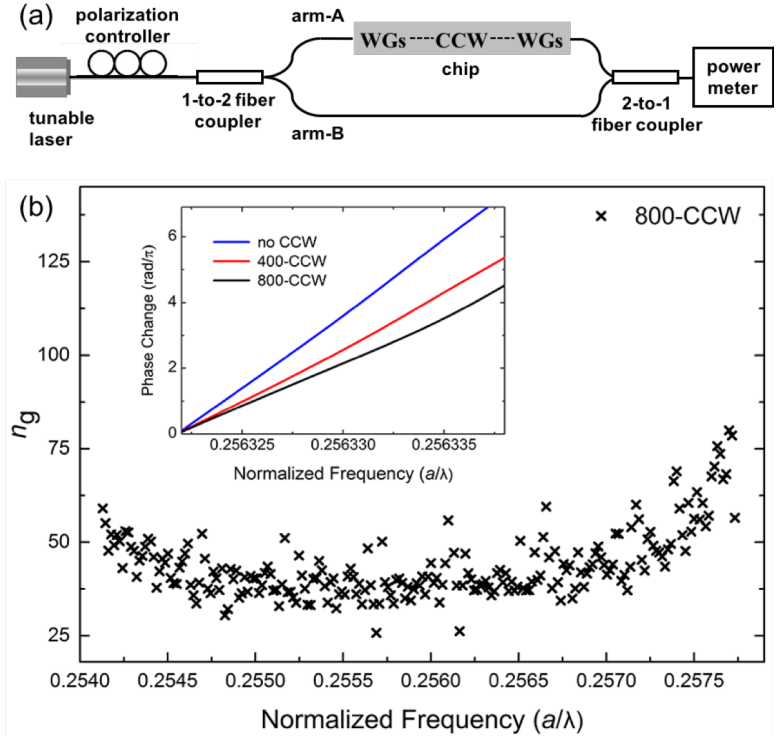

**Figure 4** | (a) Schematic of the optical fiber-based Mach-Zehnder interferometer set up. Each CCW was singly coupled to a left and right on-chip waveguide system (WG). (b) Calculated $n_g$ when a 800-CCW was present in arm-A. (Inset) Change in the relative phase, $\Delta\varphi(\omega) - \Delta\varphi(\omega_0)$, with respect to $\omega_0 = 2\pi c/(1560.54 \text{ nm})$, as obtained from the MZ fringes when in arm-A there was no CCW (blue), a 400-CCW (red), and a 800-CCW (black). The slope of the curves decreased when longer CCWs were inserted as the arm length difference decreased.

To confirm the FSI results, we carried out an alternative experimental investigation based on a MZ interferometer[24] (Fig. 4a) consisting of a CCW in one arm (arm-A) and an external optical fiber in the second arm (arm-B). (See also Supplementary Section S4.) Fig. 4b (inset) shows the change in the relative phase for three configurations of the MZ. Fig. 4b shows the calculated group indices when an 800-CCW was inserted in the arm-A. The results are in excellent agreement with the $n_g$ measured via FSI. The two techniques are complementary, with the MZ method more suited for measuring longer CCWs (given that the uncertainty in ng scales with $1/L_c$) and the FSI more suited to local investigations of the operating structure.

Together with setting a new record in the GBP of PCN-based CCWs, the nanophotonic structures reported in this letter have many potential applications as building blocks in slow-light-based devices. Particularly attractive is the implementation of slow-light-enhanced spectroscopic interferometers[25]. In such devices, the resolution can be increased by a factor as large as $n_g$ and, as demonstrated here, this performance can be extended over a broad bandwidth. High resolution spectrometers integrated in chip-scale platforms can find transformative applications in chemical and biosensing.

## Methods

**Material and electron-beam lithography**. The silicon-on-insulator wafer consisted of a 220 nm top silicon layer and a 3 μm buffer oxide layer on a silicon substrate. The CCW pattern was defined by 100 kV electron beam lithography direct writing using positive e-beam resist. The pattern was transferred from the resist (ZEP-520A) to the silicon top-layer by fluorine based inductively coupled plasma dry etching. To undercut the buried oxide layer, we used wet buffered oxide etchant while protecting the spot size converter by photo-resist. Engineered lateral openings in the membrane made the suspended structures free from buckling.

## Acknowledgments

Y. L. and A. B. acknowledge partial support from National Science Foundation (NSF) under CAREER Grant No. ECCS- 1454021. Fabrication was partially performed at the Cornell NanoScale Science & Technology Facility. M. M., V. S., M. S. M, and R. H. acknowledge funding from the Swiss National Science Foundation through projects 200021_146998, 200020_162657, 200020_149537 and 200020_169590.

# Supplementary Information

## S1. CCW input-output coupling

An end-fire design was implemented to couple light into and out of our CCW[1]. A continuous wave (CW) laser with tuning range from 1480 nm to 1660 nm was butt-coupled to the chip via a tapered and lensed optical fiber. A spot-size converter, comprising a SU-8 polymer waveguide with mode-field diameter matched to the lensed fiber, was used to collect the light from the lensed fiber (Fig. S1a). The coupled light was then adiabatically transferred from the SU-8 waveguide to the Si strip waveguide via an inverse taper waveguide. A strip waveguide-PC waveguide coupler was used to couple the light into the PC waveguide and the CCW (Fig. S1b). The transmission spectrum of the CCW was collected likewise by a different lensed fiber at the opposite facet.

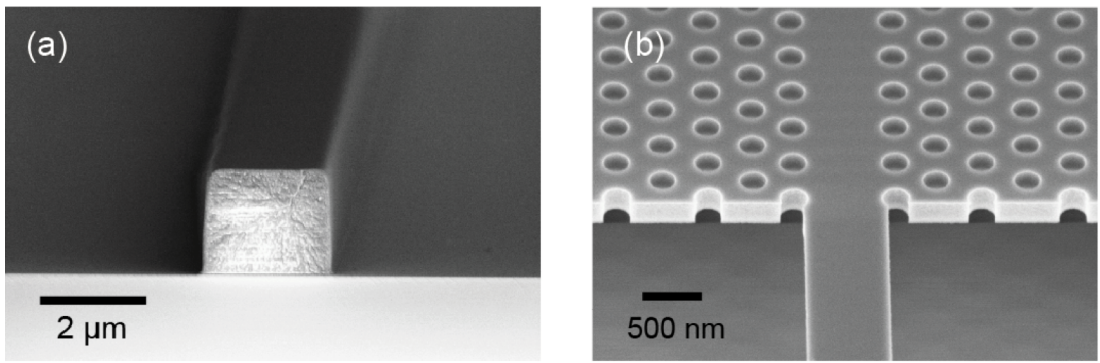

**Figure S1** | Cross-section SEM images of the (a) SU8 waveguide and (b) the interface between the Si strip waveguide and the PC waveguide.

## S2. Fourier-space Imaging

The Fourier-space Imaging (FSI)[2] was performed first by collecting the radiation of the CCW positioned at the focal plane of a high numerical aperture (NA) objective. A spatial filter was applied at the conjugate image plane to select the region of interest along the CCW and to suppress stray light, before transforming the collected light by a lens that focused the light onto a CCD detector to capture the Fourier plane. The experimental acquisition showed in Fig. 3 displayed the dispersion relation of both forward and backward propagating modes of the CCW, appearing mirrored about the symmetry axis ($k_y = 0$) and folded at the first Brillouin zone boundary ($k_y = \pm \pi/2L_y$). The measured intensity was in proportion to their respective radiative contribution, given that the CCW modes lie above the light line. The forward propagating mode was clearly dominating the transport channel, especially in the central constant $n_g$ range of the bandwidth, where the CCW operated in the dispersive regime[3].

A quasi-continuous dispersion relation could be traced through the experimental data, which was dictated by the collective spectral response of the constituent cavities. The inherent discretization of states due to the finite cavity number was apparent especially with the shorter CCWs, since the CCW design utilized a broad bandwidth span relative to the linewidth of the formed states. The dispersion profile smoothed when the CCW chain was elongated, with a proportional number of states being created, which narrowed down the inter-state frequency gaps.

Moving towards the edge of the band, $n_g$ gradually rose and scattering into the backward-propagating mode became progressively more significant as a consequence of disorder-induced scattering. When $n_g$ exceeded a value of around 50, the dispersion linewidth, which is naturally governed by the finite-sample size and attenuation, began to broaden, signaling the onset of diffusive light transport and subsequent localization in the photonic crystal lattice towards higher $n_g$ values. Localized states exhibited spectral signatures with a broad extent in $k$-space and corresponding dips in light transmission.

**S3. Additional series of CCW devices**

To explore the space of parameters around the GBP optimal design, three additional CCW designs targeting higher $n_g$ were fabricated and measured. Table S1 compares the experimental results for all four designs. D1 is the design described in the article, which exhibits the highest GBP. Fig S2 shows an example of the measured FSI and normalized transmission for a D4 design with a CCW formed by 50 coupled PCCs. Our measurements showed systematically lower GBPs and lower transmission bandwidths for D2-4. This was an expected result because a higher $n_g$ implies a higher susceptibility to fabrication imperfections. We emphasize, however, that D2 represents still one of the PCN-based CCW slow light devices with largest bandwidth at such high $n_g$.

|    | $\langle n_g \rangle$ | $\Delta\lambda$ | GBP  |
|----|------|------|------|
| D4 | 47.0 | 12.4 | 0.38 |
| D3 | 81.2 | 5.8  | 0.30 |
| D2 | 94.5 | 3.1  | 0.19 |
| D1 | 41.0 | 17.7 | 0.47 |

**Table S1** | Average index of refraction ($\langle n_g \rangle$) and bandwidth ($\Delta\lambda$) measured for three additional series of CCW devices. D1 corresponds to the device reported in the article.

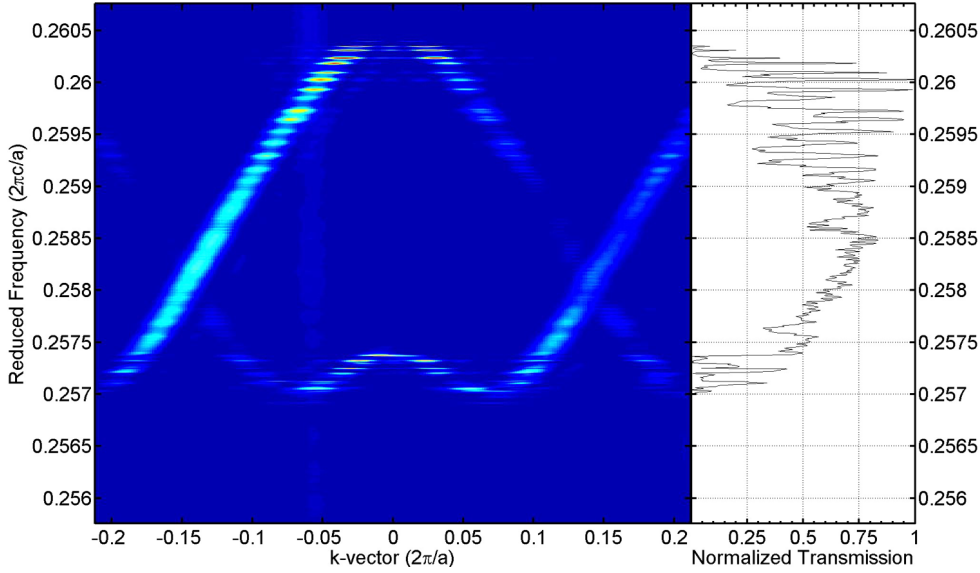

**Figure S2** | (a) Photonic band structure measured by FSI and normalized transmission of a D4 design CCWs formed by 50 PCCs (50-CCW).

**S4. CCW-based Mach-Zehnder interferometry**

Interference in our CCW-based Mach-Zehnder (MZ) (Fig. 4a) was measured as a function of the input laser wavelength. When in arm-A the two tapered fibers were directly coupled through a small air-gap (i.e., the chip was entirely removed), the MZ fringes (Fig. S3b) showed the optical path in the arm-B to be ~ 6 cm longer than in the arm-A. We defined the relative phase between the two arms without chip as $\varphi_{r0} = \varphi_B - \varphi_A$ and with the chip as $\varphi_r = \varphi_{r0} - \omega\,(n_c L_c + n_w L_w)/c$, where we defined a CCW with effective index $n_c$ and length $L_c$, and the chip-integrated input-output coupling waveguides (indicated as WGs in the schematics of Fig. S3a) with effective index $n_w$ and total length $L_w$. As we scanned the laser frequency, the change in the relative phase was $d\varphi_r/d\omega = d\varphi_{r0}/d\omega - (n_g L_c + n_{gw} L_w)/c$, where $n_g$ and $n_{gw}$ are the group indices of the CCW and the coupling waveguides. From the chip design and the MZ fringes observed in presence of the different CCWs, we calculated that the entire left and right couplings waveguide system contributed for about 1.2 cm to the optical path length of the arm-A. The change in the relative phase, $\Delta\varphi(\omega) - \Delta\varphi(\omega_0)$, for three configurations of the MZ are displayed in Fig. 4 (inset). The group index of the CCW was obtained as

$$n_g = \frac{c}{L_c}\left(\frac{d\varphi_{r0}}{d\omega} - \frac{d\varphi_r}{d\omega} - \frac{n_{gw}L_w}{c}\right)$$

with $d\varphi_{r0}/d\omega$ and $d\varphi_r/d\omega$ measured by the MZ fringes. Because the error of the CCW group index ($\delta n_{gc}$) is related to the error in the measurements as $\delta n_{gc} = (c/L_c)\,\delta[d\varphi_{r0}/d\omega - d\varphi_r/d\omega]$ and is proportional to $1/L_c$, the interferometry method turned out to be better suited than the FSI to measure longer CCWs.

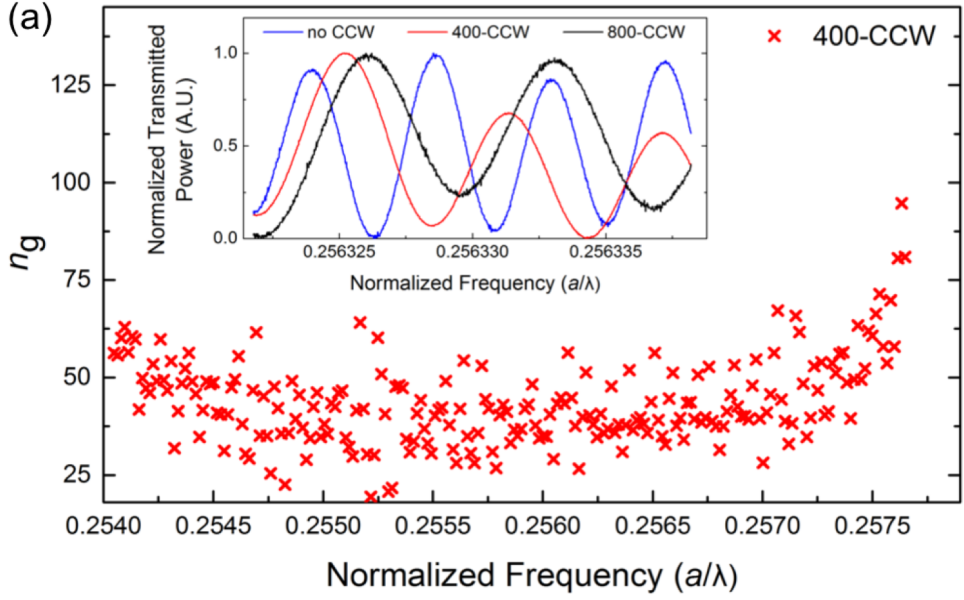

**Figure S3** | (a) Calculated $n_g$ for 400-CCW. Inset: Normalized transmitted power for $\lambda \sim 1560.54$ nm when in arm-A there was no CCW (blue), 400-CCW (red), and 800-CCW (black).